\documentstyle[11pt]{article}
\newcommand \be  {\begin{equation}}
\newcommand \bea {\begin{eqnarray} \nonumber }
\newcommand \ee  {\end{equation}}
\newcommand \eea {\end{eqnarray}}

\begin{document}

\begin{center}
\centering{\bf \Large Gauge theory of Finance?}
\end{center}

\begin{center}
\centering{Didier Sornette$^{1,2}$\\
{\it $^1$ Department of Earth and Space Science\\ and Institute of Geophysics
and Planetary Physics\\ University of California, Los Angeles,
California 90095\\
$^2$ Laboratoire de
Physique de la Mati\`ere Condens\'ee, CNRS UMR6622\\ Universit\'e des
Sciences, B.P. 70, Parc Valrose, 06108 Nice Cedex 2, France
}
}
\end{center}

\vskip 1cm
{\bf Abstract\,:} {\small Some problems with the recent stimulating proposal of
a ``Gauge Theory of Finance'' are outlined.
First, the derivation of the log-normal distribution
is shown equivalent both in information and
mathematical content to the simpler and well-known derivation, dating back 
from Bachelier and Samuelson. 
Similarly, the re-derivation of
Black-Scholes equation is shown equivalent to the standard one because the limit of no uncertainty
is equivalent to the standard risk-free replication argument. Both
re-derivations of the log-normality and Black-Scholes result do not provide a test of
the theory because it is degenerate in the limits where these results apply. Third,
the choice of the
exponential form a la Boltzmann, of the weight of a given market configuration, is 
a key postulate that requires justification.
In addition, the ``Gauge Theory of Finance'' seems to lead to ``virtual''
arbitrage opportunities for pure Markov random walk market when there should be none.
These remarks are offered in the hope to improve the formulation  of
the ``Gauge Theory of Finance'' into a coherent and useful framework.}

\vskip 2cm

Ilinski and collaborators  \cite{IL1,IL2} have recently proposed an intringuing analogy between
quantum electrodynamics (QED), gauge field theory and financial markets (see also
\cite{Dunbar} for a simplified introduction).
Their main idea starts from the recognition that
 the quest for arbitrage \footnote{Arbitrage, also known as the Law of One Price, states that two
assets with identical attributes should sell for the same price and so should the same asset
trading in two different markets. If the prices differ, a profitable opportunity arises
to sell the asset where it is overpriced and to buy it where it is underpriced.}
opportunities by market players is one of the possible reasonable
underlying explanation for anomalous behaviors such as deviations from log-normality
of price variations. 
They then proceed to develop a geometrical theory in which an
arbitrage opportunity is equivalent to a local curvature of the plaquettes corresponding
to the different financial intruments. This is also very intuitive since a parameter-free
definition of a (non-zero) curvature involves the non-closure of 
a parallel transport along a closed
path, which is quite similar to an investment from its inception to closure. In other
words, it is nothing but the standard discounted procedure.
They then construct a set of axioms and then claim to provide an 
``intuitive'' explaination for the log-normal distribution reference 
of price variations and to have the potential 
for a systematic analysis of deviations from the Black-Scholes option equation.

Here, I first show that their derivation of the log-normal distribution
is in fact equivalent both in information and
mathematical content to a simpler and well-known derivation, essentially dating back 
from Bachelier \cite{Bachelier}. This is perhaps somewhat hidden behind the gauge field
formulation (at least for non-physicists) and thus worth stressing. 
Similarly, their re-derivation of
Black-Scholes equation is equivalent to the standard one because the limit of no uncertainty
is equivalent to the standard risk-free replication argument \cite{BS}.
My message is that both
re-derivations of the log-normality and Black-Scholes result do not provide a test of
the theory because it is degenerate in the limits where these results apply. In particular,
the key postulate of Ilinski and collaborators \cite{IL1,IL2}, namely the 
exponential form a la Boltzmann of the weight of a given market configuration, is in no way
justified a priori. It is convenient for calculations but its validity must be tested 
thoroughly.  Physical analogies are useful but must often be adapted to 
the market constraints and their specificities. For physical systems, 
the Boltzmann weight form has a strong
mathematical basis derived from equally strong physical constraints for physical systems.
Why should the same
considerations work for the market? Samuelson, when referring to the use of entropy
in economics, said ``... I have come over the years to have some impatience and
boredom with those who try to find an analogue of the entropy of Clausius or Boltzmann
or Shannon to put into economic theory'' \cite{Samuelson}. A further argument is that
the Boltzmann statistics was generalized in the last ten years into the Tsallis
statistics \cite{Tsallis}, suggesting that not all phenomena in nature
have to be described by Boltzmann statistics. 
Furthermore, the gauge theory of arbitrage is by construction 
such that the presence of noise introduces ``virtual arbitrage'' opportunities
\cite{IL1,IL2}.  However, take
a pure Markov random walk market\,: by construction, there are no arbitrage opportunities but
a lot of noise. 

Consider buying a security at price $S_1$ and selling it at price $S_2$ a time $\tau$ later.
The return is ${S_2 \over S_1} - 1$. Suppose that instead you had sold at price 
$S_1$ and bought later (thus closing your position) at price $S_2$ a time $\tau$. 
The return is now ${S_1 \over S_2} - 1$. Not knowing the direction of the security price, 
Ilinski and collaborators \cite{IL1,IL2} argue that it is natural to form a
strategy where you buy and sell simultaneously. Then the return is 
${S_1 \over S_2}+ {S_2 \over S_1} - 2$. In the limit of small time increment, the return
per unit time is \footnote{This portfolio gives a certain return! But it is very small.
With a typical daily price variation of about $1\%$, the return of this
``long$+$sell'' strategy is $0.005\%$ per day or $1.25 \%$ per year.}
\be
{1 \over \tau} [{S_1 \over S_2}+ {S_2 \over S_1} - 2] \approx {\tau \over 2} 
({\delta \ln S \over \delta \tau})^2~.
\ee
Plugging this expression in the postulated exponential probability for such an event to occur,
we get the Wiener integral \cite{Wiener} whose solution is well-known to be the Gaussian distribution, 
hence the log-normal distribution for the log-price.
In addition, the Wiener integral is the mathematical integral description
of the random walk process. This is at the core of the Edwards'model of polymers
\cite{Edwards}. This is the essence of the ``intuitive'' derivation of 
Ilinski and collaborators \cite{IL1}. Superficially, this seems to substantiate the choice of the 
exponential form of the probability, here in question. But this justification is based
on the specific choice of the constructed portfolio. There are many different securities
with complex non-linear dependence on underlying stocks and it is not clear at all what
warrants this approach. One could argue that the Boltzmann weight assumption
is not  restrictive until one does specify the action. Basically,
the question is about the action.
Ilinski and collaborators  \cite{IL1,IL2} follow standard arguments for small
fluctuations and propose a quadratic action. The problem is that the quadratic form
depends on the exponential Boltzamnn weight assumption and on the choice of the 
portfolio. For instance, consider the choice of the magnetization $M$ as the natural variable
to describe a magnetic material. It leads to 
the Landau potential with a leading quadratic term. In constrast, suppose an alien was to choose
$\log M$ as the natural variable. Then, the theory in terms of $\log M$ becomes
highly involved with a multifractal structure.

Let us now examine the standard alternative derivation. Make an investment where you buy
a security at price $S_1$ and sell it at price $S_2$ a time $\tau$ later.
The corresponding return per unit time is ${\delta \ln S \over \delta \tau}$. Now, following
Bachelier, the absence of arbitrage opportunities corresponds to assume that 
${\delta \ln S \over \delta \tau}$ is random, with no memory\,:
\be
 {\delta \ln S \over \delta \tau} = \eta(\tau)~,
 \label{redsz}
 \ee
 where $\eta$ is a white noise with an arbitrary distribution. 
 This is nothing but the equation of a random walk.
 By the central limit theorem, we retrieve the usual log-normal result.
Expression (\ref{redsz}) is the stochastic differential equation solution of the Wiener integral,
thus showing the link between the two approaches. There is thus nothing particular in the 
derivation of Ilinski and collaborators \cite{IL1}. A completely similar
correspondence holds for their derivation \cite{IL2} of the Black and Scholes option 
pricing equation. Their idea
here is to construct the relevant portfolio and transaction steps, then take the 
continuous time limit and go to the limit of no uncertainty which amounts to take the 
stationary solution of the exponential weight. The limit of no uncertainty is nothing 
but the standard replication argument of Black and Scholes \cite{Hull}
for a complete market.

In conclusion, the idea of Ilinski and collaborators \cite{IL1,IL2}
is potentially very interesting. However, the results on the log-normal distribution and
the Black and Scholes equations are not proof of the correctness of the theory or even 
of its relevance. 
The same can be said for other general theories\,: general relativity is not correct because
it reduces to Newton gravitation at small energies. This is only a requirement of
correspondence (this point is indeed emphasized in Ref.\cite{IL1,IL2}). Other
principles must apply to justify its construction. The concept of arbitrage 
is probably correctly advanced as a fundamental ingredient of the theory. The problem is that 
the theories that incorporate this ingredient are non unique. In fact, the finance 
litterature is all about arbitrage. Ross \cite{Ross} wrote probably the first paper
concerned with the essential role of the no-arbitrage condition in valuation. This led
afterwards to the development of the general theory of asset pricing in terms of 
a valuation operator (the so-called stochastic discount factor) \cite{Cochrane}. 
The problem, that any
theory must face, is that for real (incomplete) markets, the valuation operator is not unique.
The fundamental question is\,: which valuation operator should be used? 
How then does the gauge formulation
of finance help shed light on this problem?

\vskip 0.5cm

I acknowledge stimulating exchanges with K. Ilinsky, 
L. Martellini, D. Stauffer and B. Urosevic.

\end{document}